\documentstyle[epsf,aps,prl]{revtex}

\begin{document}
\draft
\twocolumn[\hsize\textwidth\columnwidth\hsize\csname
@twocolumnfalse\endcsname
\title{ ``Superconductor-Insulator transition'' in a single Josephson
junction }
\author{J.S. Penttil\"a$^1$, \"U. Parts$^1$, P.J. Hakonen$^1$, M.A.
Paalanen$^1$,
and E.B. Sonin$^{1,2}$}
\address{ $^1$Low Temperature Laboratory, Helsinki University of Technology,
FIN-02015 HUT, Finland \\
$^2$The Racah Institute of Physics, The Hebrew University of Jerusalem,
Jerusalem 91904, Israel}

\date{\today} \maketitle

\begin{abstract} $VI$-curves of resistively shunted single Josephson junctions
with different capacitances and tunneling resistances are found to display a
crossover between two types of $VI$-curves: one without and another with a
resistance bump (negative  second derivative) at zero-bias. The crossover
corresponds to {\em the dissipative phase transition} (superconductor-insulator
transition) at which macroscopic quantum tunneling  delocalizes the Josephson
phase and destroys superconductivity. Our measured phase diagram does not agree
with the diagram predicted by the original theory, but does coincide with a
theory that takes into account the accuracy of voltage measurements and
thermal
fluctuations.

\end{abstract}
 \bigskip
]

A Josephson junction is an unique physical object on which one can test a great
variety of important physical concepts of modern physics, such as
macroscopic quantum tunneling of the phase,
quantum mechanical coherence, Coulomb blockade
etc. An important place in this list is occupied by the so called
{\em dissipative phase transition} (DPT), predicted for various systems
\cite{Letal,SZ,W}.
The physical origin of this transition is the suppression of macroscopic
quantum
tunneling of the phase by the interaction with dissipative
quantum-mechanical environment, described by the Caldeira-Leggett model.
Macroscopic quantum tunneling destroys superconductivity of a junction,
whereas  suppression of tunneling restores Josephson current. Hence, this
transition is often called a superconductor-insulator transition (SIT).

The detection of DPT in a {\em single} Josephson junction is
of principal importance since it is the simplest system where this transition
is expected, without any risk of being masked by other physical processes, as
is possible in more complicated systems like regular or random
Josephson junction arrays. Some evidence of DPT (SIT) in a
single Josephson junction has already been reported
\cite{Jap}, but only for the case of weak Josephson coupling. It has not been
enough to trace the whole phase diagram, including the range of strong
Josephson coupling where the theoretical predictions are especially
intriguing.

In this Letter we present results of our measurements on $R=dV/dI$ vs. $I$
curves, for a variety of single small isolated Josephson junctions, shunted and
unshunted, with different values of capacitance $C$ and normal state tunneling
resistance $R_T$. We have detected a crossover between two types of RI-curves
with an essentially different behavior at small currents. Relating this
crossover with the DPT, we were able to map out the {\em whole} phase
diagram for
a Josephson junction. The position of the observed phase
boundary does not agree with that expected from the original theory. However,
the theory revised to take into account a finite accuracy of our voltage
measurements ({\it viz.}, the minimum voltage that we are able to detect),
explains well the observed phase diagram.  We also argue that the real
signature of
DPT is a modification of $VI$-curves as observed in our experiment:
the SIT,
traditionally defined as the change of sign of  
thermoresistance $dR/dT$, is
not necessarily identical to the DPT. The measured phase diagram provides the
first observation of DPT for a single Josephson junction in the whole interval
of the Josephson coupling.

\begin{figure}[tb]
\begin{center}
\epsfxsize=80mm
\epsffile[48 398 541 595]{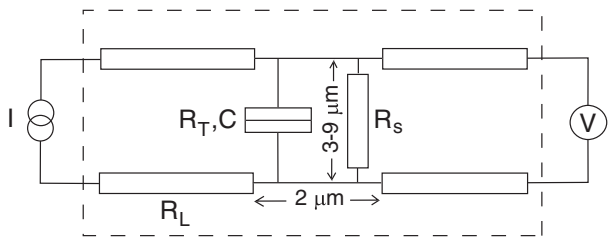}
\caption{Schematic diagram of shunted tunnel junction in a resistive
environment.}\label{f1}
\end{center}
\end{figure}

Our sample consists of a shunted superconducting  Al-AlO$_x$-Al tunnel
junction
(area 150*150 nm$^2$). Its resistance  $R_T$ = 3.4 - 21 k$\Omega$ was
determined by reducing the shunt resistance $R_s$ = 4 - 75 k$\Omega$
off from 
the normal state resistance measured
at 0.1 K. The shunted junction was connected to four measurement
leads via 20 $\mu $m long thin film Cr resistors
$R_L \approx$ 100 k$\Omega$. This ensures
a well-defined resistive environment governed by the shunt (see Fig.~\ref{f1}).
The value of $R_s$ was deduced using the length
of the shunt and the measured resistivity of the Cr sections in the leads.
The circuits, both shunted and unshunted, were fabricated
using electron beam lithography and triple-angle evaporation.
The Cr resistors and shunt (10-15 nm thick, 100 nm wide) were
evaporated
at right angle of incidence. When exposing  
the chrome metal sections in e-beam writing, an accurately tuned electron dose 
ensured that the Al replicas were evaporated on the side of the resist
and thus removed during lift-off.
Within 5 $\%$, no change was observed in $R_s$ when $B$ and $T$
were swept over 0 ... 0.2 T and 0.1 ... 4 K, respectively. On the dilution
refrigerator, the samples were mounted inside a tight copper enclosure and
the measurement leads were filtered
using 0.5 m of Thermocoax cable.

Two types of observed RI-curves  are shown in Fig.~2. In the
``superconductor''-type (Fig.~2a), the resistance
has its minimum at zero bias and increases monotonically
up to sub-gap resistance (in parallel with $R_s$) given by the maxima in
the figures. In the ``blockade'' type (Fig.~2b), a higher resistance
``bump'' appears at small currents, i.e., the resistance is maximum at zero
bias. The width of this feature becomes more pronounced with decreasing
Josephson coupling \cite{Hav}. In both cases, the resistance returns
smoothly to its normal state value
after the subgap maximum.  

In order to determine the phase diagram, we plot in Fig.~\ref{f5}
the character of our samples 
(``superconductor''/''insulator'') on the coordinate plane
 ($R_q/R$, $E_J$/$E_C$) which are the intrinsic
parameters of DPT \cite{SZ}.  The DPT boundary is to
separate open and solid symbols in Fig.~\ref{f5}. 
Here $R_q=\frac{h}{(2e)^2}=6.5$
k$\Omega$ is the
quantum resistance, and the resistance $R=R_s R_{qp}/(R_s +R_{qp})$
characterizes the total ohmic dissipation, $R_{qp}$ being the quasiparticle
resistance.
The Josephson energy $E_J$ was calculated from $R_T$
using the Ambegaokar-Baratoff relation while
the Coulomb energy $E_C=e^2/2C$ was estimated from normal
state $VI$-curves: the junction capacitance was obtained
from the offset at large bias voltages using the formula
$V_{\mathrm{offset}}=\frac{e}{2C} \frac{R_s}{R_s+R_T}$ that
comes from a simple balance of currents through the junction and the shunt.
As seen from the summary of junction parameters in Table I, the
ratio $E_J$/$E_C$ falls between 0.85 and 14.1.

Zero bias resistance $R_0$ is displayed as function of temperature in 
Fig.~\ref{f3} .
Near $T_c$ all the samples have a pronounced peak which has also been
observed by Shimazu {\it et al} \cite{Shi}. At lower temperatures, the
resistance first
decreases and then starts to increase again (Fig. 4b).
This re-entrant behavior is not observed in samples with
large $E_J/E_C$ which stay ``superconducting'' all the way down to lowest
temperatures (Fig.~4a).

Let us discuss the phase diagram for DPT expected from theory. In the
classical limit at zero temperature, the Josephson phase $\varphi$ is
trapped in
some well of the ``tilted washboard'' potential $U(\varphi) =- E_J\cos
\varphi -(\hbar/2e)I\varphi$. This localized-phase state corresponds to a
superconducting state. In fact, this localization is never perfect: (i) at
finite
temperatures the phase can hop from one well to another via thermal activation;
(ii) at very low temperatures the phase is able to escape from a well via
macroscopic quantum tunneling which is an exponential function of the barrier
height
$\propto E_J$ \cite{T}. When we say that the
junction is a ``superconductor'' we mean that (i) its resistance is
essentially smaller than the normal junction resistance, (ii) its resistance
increases with temperature like in a metal because enhanced thermal
fluctuations
produce an increased phase slip rate $d\varphi/dt$.

Because of quantum-mechanical tunneling, the bound states in
different wells form an energy band like in a solid \cite{LZ}. The band energy
is a periodic function of the quasicharge $Q$ (an analog of quasimomentum in a
solid) with the period $2e$. If $E_J/E_C \gg
1$, that corresponds to the ``tight binding'' limit in the solid-state
theory, then $E(Q)=E(Q+2e)= \Delta [1 -\cos (\pi {Q/e})]$.
Here the band half-width is given by \cite{LZ}
\begin{equation}
\Delta={16 \over  \sqrt{8\pi}}\left(E_J \over 2E_C\right)^{1/4} \hbar \omega_p
\exp\left[ - \left(8 E_J \over E_C \right)^{1/2}\right] ,
   \label{Band} \end{equation}
where $\omega_p=\sqrt{8E_J E_C}/\hbar$ is the plasma frequency.
For a small
quasicharge
$Q\ll 2e$ one may use the ``effective-mass'' approximation $E(Q)={Q^2 / 2C^*}$
where the effective capacitance (an analog of the effective mass) $C^*={e^2 /
\pi ^2 \Delta} $ can exceed the geometric capacitance $C$ essentially.

The band theory predicts  Ohm's law $V=RI$ at small current bias $I \ll
e/RC^*$. This corresponds to the quasicharge $Q=C^* V=IRC^*$.
However, with increasing current bias the quasicharge approaches the
Brillouin-zone boundary ($Q=\pm e$). Then another regime of phase motion sets
in \cite{LZ}:  The phase performs Bloch oscillations, $\varphi=2e E(Q)/
\hbar
I $, with $Q=It$ leading to the period $2e/I$.
In this regime dissipation is suppressed,
corresponding to a decreasing resistance $V/I$.

Thus, at small current bias the RI-curve must have a
bump of width $e/RC^*$ (a voltage of $e/C^*$), and the Josephson
junction {\em always} behaves as a normal junction with ohmic resistance $R$.
At larger currents $I \gg e/RC^*$, however, the junction has a tendency to
become superconducting again. This behavior is a direct outcome of
the band picture for the phase motion, as was shown in Ref. \cite{LZ}.
It was obtained also using more rigorous path-integral
methods
\cite{SZ,W}. Therefore, a blockade bump in the RI-curve of a Josephson
junction is a clear
manifestation of phase delocalization and the band picture.

The bump on the RI-curve at small bias looks similar to the bump due to the
Coulomb blockade of  single-electron tunneling and, moreover, is governed
by the
same effective Coulomb energy $e^2/C^*$. On the other hand,
in the model which we are discussing here, there is no single-electron
tunneling
at all if the resistance $R$ is dominated by the shunt resistance $R_s$
(quasiparticle resistance $R_{qp} >> R_s$).  In fact, we
deal with the Coulomb blockade indeed, but it is the Cooper-pair current
channel that is blocked \cite{Hav}. However, in an unshunted junction with $R =
R_{qp}$ the additional Coulomb blockade
of single-electron tunneling can increase the zero-bias
resistance well above $R$.

The theory as summarized above would indicate that any Josephson
junction must have a blockade bump at zero bias. However, we must take into
account an important effect of the environment: suppression of the quantum
tunneling between wells by dissipation
\cite{Letal,SZ,W}. This decreases the band half-width which now is given by
\begin{equation}
\tilde \Delta = \Delta\left(\Delta \over\hbar \omega_p
\right)^{\alpha \over 1- \alpha} .
 \label{Tband} \end{equation}
Here $\alpha=R_q/R$ is the dissipation parameter. The renormalized
energy $\tilde \Delta$ vanishes at $\alpha=1$ where the band
disappears and quantum tunneling becomes impossible \cite{QT}. Then, the
junction is superconducting down to the lowest current bias. Consequently,
the phase line separating the ``insulator'' from the ``superconductor'' is
the $\alpha =1$ line independently of the energy ratio $E_J/E_C$ (the dashed
vertical line on
the phase diagram, Fig.~3)
\cite{PD}.

This phase diagram, in which the Josephson junction under weak 
dissipation remains an
``insulator'' even in the limit of $E_J/E_C \rightarrow \infty$, is difficult
to confirm because the putative, very slow delocalization of phase leads to 
exceedingly small voltages.
Experimentally, the ``insulator'' behavior can be observed only if the
voltage of the bump, the effective Coulomb gap $e/C^* \sim \tilde \Delta/e$,
exceeds the minimum voltage $V_{min}$ detectable in our measurements.
Therefore,
it is reasonable to assume that our measured DPT corresponds not to the
condition $\tilde \Delta =0$, but to $ \tilde \Delta  \sim e V_{min}$. Together
with Eqs. (\ref{Band}) and (\ref{Tband}) (neglecting an unimportant factor of
$(E_J/2E_C)^{1/4}$ in Eq. (\ref{Band})),  the latter condition yields the
crossover from the superconductor to the insulator behavior at
\begin{equation}
{E_J \over E_C}= {1 \over 8} \left(\ln{16 \over \sqrt{8\pi}} +(1-\alpha)\ln
\omega_p \tau_s \right)^2~.
   \label{alpha} \end{equation}
Here  $\tau_s =\hbar/eV_{min}$ is the  phase slip time for the
minimum detectable voltage $V_{min}$. This is the time necessary for a
phase change
by $2\pi$, {\it i.e.}, for the phase motion between two wells.  In our
case, $V_{min}$ is
about 0.5 $\mu$V which corresponds to $\tau_s \approx 2\cdot 10^{-9}$ s. The
curve obtained from Eq.~(\ref{alpha}) using the plasma
frequency $2\cdot 10^{11}$ Hz is displayed in Fig.~\ref{f5}.
Within our quite large
statistical uncertainty, Eq.~(\ref{alpha}) agrees
with the experimental crossover between
"superconductor" and "blockade" types of $RI$-curves.
If we apply the argument by Sch\"on and Zaikin \cite{SZ} 
that an insulator state is observable when the
phase spreading time  $\hbar /\tilde \Delta$
is smaller than the observation time $\tau$ in our
experiment, then
the crossover (replacing $\tau_s$ by $\tau \sim $ 1 s in Eq.~(\ref{alpha})) 
would take place at $E_J/E_C \approx 100$ in contrast to
$E_J/E_C \sim 10$ observed in the experiment.
Thus, the ability to reveal the blockade bump (insulator behavior) is
restricted
not by the observation time, but by the accuracy of the voltage measurement.

According to Ref.~\cite{LZ}, thermal fluctuations are also able to ``wash out''
the blockade bump if thermal energy $kT$ is
on the order of or larger than $e^2/C^*$. In this case, the
crossover is given by Eq.~(\ref{alpha}) again, but with $ \tau_s$ replaced by $
\hbar/kT$ which is about 5 times less
than $ \tau_s$ at our minimum temperature of 50 mK. Since the crossover
depends logarithmically on $\tau_s$, small uncertainties in $\tau_s$
do not shift its position essentially when compared with our experimental
uncertainty.
In fact, since the numerical factors in the conditions $kT \sim
e^2/C^*$ and
$V_{min} \sim e/C^*$ are not known, it is difficult to judge which one of these
restrictions is stronger.

Finally, we want to compare the concepts of {\em the superconductor-insulator
transition} (SIT) and {\em the dissipative phase transition} (DPT). The common
formulation is that  ``superconductor'' and
``insulator'' are specified by the positive and negative sign of $dR_0/dT$,
respectively. Accordingly, one may identify the peak in $R_0 (T)$ 
(see Fig.~\ref{f3})
also as SIT. But the SIT near $T_c$ has nothing to do
with DPT predicted theoretically \cite{Letal,SZ,W}: the peak in Fig.~\ref{f3}
corresponds to the temperature at which the normal junction (with the 
$RI$-curve noted by $N$ in Fig.~\ref{f5}) 
becomes superconducting, i.e., a Josephson
junction with a {\em detectable} critical current ( 
$RI$-curve noted by $S$ in Fig.~\ref{f5}). The
DPT theory assumes that the critical current is initially finite, but in
reality
it may be essentially smaller than $eE_J/\hbar$ because fluctuations are
especially important at small $E_J/E_C$. We believe that this
discussion is relevant for
understanding the data of Yagi {\em et al.} \cite{Jap}, who observed the
SIT for $E_J/E_C$ between 0.1 and 0.2
for strong dissipation $\alpha >1$ (horizontal dotted line
in Fig.~\ref{f5}). These results were considered to be
contradictory to the DPT theory which does
not predict any transition to the insulating phase at  $\alpha >1$. In
fact, there is no disagreement: Yagi  {\em et al.} did not observe DPT
where the
energy band width is to vanish, but SIT where the critical current disappears.
So, our important conclusion is that the concepts of DPT and SIT are not
completely identical: both are accompanied by the sign change of $dR_0/dT$ and
thus any DPT is SIT, but {\em not vice versa}\cite{SIT}. In our case the zero slope of
$dR_0/dT$ occured at about $R_s=10$ k$\Omega$, which roughly agrees with
the DPT
line obtained from the RI-curves. 

In summary, our experiments clearly confirm the existence of the
dissipative phase transition in a single Josephson junction, though the
observed phase diagram is quite different from that expected originally. The
agreement with theory is achieved by taking into account that the
position of the measured phase boundary is governed not only by 
intrinsic junction parameters, but also by the accuracy of voltage
measurements. Our work is a strong demonstration of quantum effects in a
single Josephson junction, especially, of the Josephson phase delocalization
and the band picture of phase motion.

We acknowledge interesting discussions with G. Sch\"on, A.D.
Zaikin, and A.B. Zorin. This work was supported by the Academy of Finland
and by the Human Capital and Mobility Program ULTI of the European Union.


\begin{figure}[tb]
\begin{center}
\epsfxsize=80mm
\epsffile[34 383 575 702]{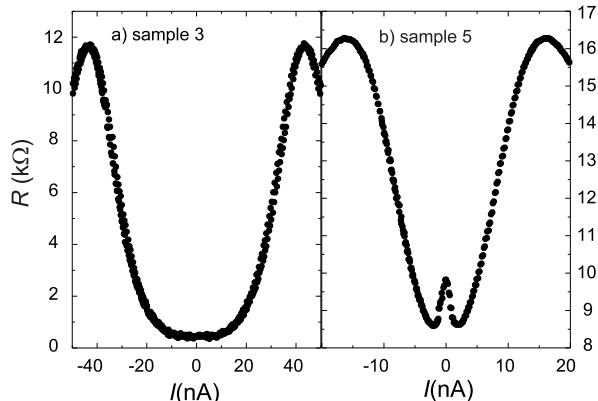}
\caption{Resistance vs. current for two samples showing different
behavior: a) Sample 3 with $R_T=3.7$ k$\Omega$ and $R_s=11$ k$\Omega$,
 b) sample 5 with $R_T=12.4$ k$\Omega$ and $R_s=22$ k$\Omega$.}\label{f2}
\end{center}
\end{figure}

\begin{figure}
\begin{center}
\epsfxsize=80mm
\epsffile[31 425 444 686]{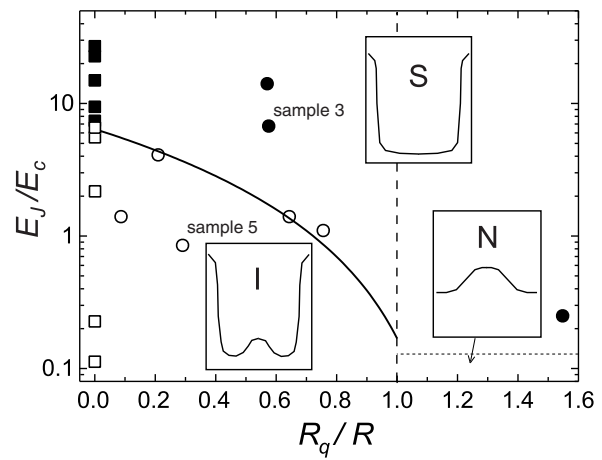}
\caption{Phase diagram of shunted Josephson junction. The phase
boundary lies between insulator-like (open symbols, I) and
superconductor-like samples (solid symbols, S). Unshunted samples
~(squares) are collected at $R_q/R = 0$.
The solid line is the theoretical phase boundary calculated
using Eq.~(3) with $\omega_p= 2\cdot 10^{11}$ 1/s and $\tau_{s}=
2\cdot 10^{-9}$ s. The dotted line is the transition line
for strong dissipation to a state with no supercurrent (N)
found by Yagi {\em et al.} [4]. For details, see text.} \label{f5}
\end{center}
\end{figure}

\begin{figure}
\begin{center}
\epsfxsize=80mm
\epsffile[70 449 507 709]{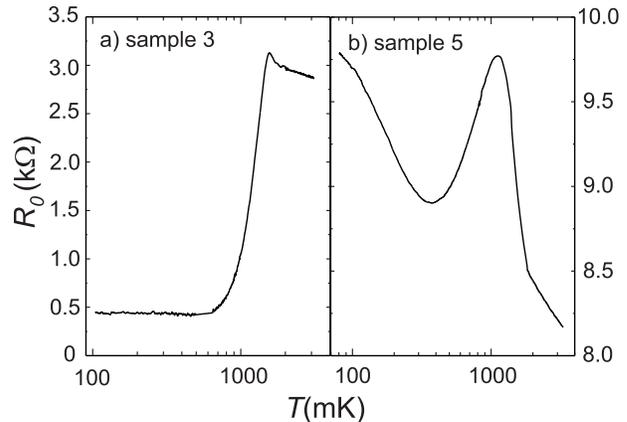}
\caption{Zero-bias resistance $R_0$ as a function of temperature for
samples 3 and 5.} \label{f3}
\end{center}
\end{figure}

\vspace{-1cm}

\begin{table}
\caption{Measured shunted junctions. $R_T$ is deduced from the slope of
the normal state IV-curve at high bias voltage.
The effect of parallel shunt $R_s$
is subtracted. $C$ is calculated from the high bias
offset voltage using $V_{\mathrm{offset}}=\frac{e}{2C} \frac{R_s}{R_s+R_T}$.
The value of $R_s$ is estimated from the known wire resistivity.}

\begin{tabular}{|@{\hspace{2mm}}c@{\hspace{2mm}}|@{\hspace{4mm}}c@{\hspace{4mm}}
|@{\hspace{4mm}}c@{\hspace{4mm}}|@{\hspace{4mm}}c@{\hspace{4mm}}|@{\hspace{4mm}}
c@{\hspace{4mm}}|}
   \hline
	sample & $R_T(k \Omega )$ & $C(fF)$ & $R_{s}(k \Omega )$ &
                 $E_J / E_c $ \\
   \hline 1 & 9.7 & 1.8  & 75 & 1.4 \\
   \hline 2 & 4.5 & 2.5  & 31 & 4.0 \\
   \hline 3 & 3.7 & 3.4 & 11 & 6.8 \\
   \hline 4 & 3.4 & 6.6 & 11 & 14.1 \\
   \hline 5 & 12.4 & 1.5 & 22 & 0.85 \\
   \hline 6 & 8.1 & 1.7 & 10 & 1.4 \\
   \hline 7 & 5.9 & 2.2 & 8.6 & 1.1 \\
   \hline 8 & 21 & 0.8 & 4.2 & 0.25 \\
      \hline
  \end{tabular}
\end{table}

\begin{references}
\bibitem{QT} Because of dissipation, the energy is not conserved
during a quantum mechanical transition. The energy loss
may be estimated as $\Delta E_d \sim (\hbar/2e)^2 (1/R \tau_{tr})$ where
$\tau_{tr}\sim (\hbar/2e) \sqrt{C/E_J} \sim 1/\omega_p$ denotes
the time of tunneling between two potential wells (the traversal time).
It is interesting to note that complete suppression
of macroscopic quantum tunneling at  $\alpha >1$ corresponds to the
condition that the
dissipated energy $\Delta E_d$ exceeds the ground-state energy $E_0 \sim \hbar
\omega_p$ in the well.
\bibitem{PD} In fact, the phase transition line depends on $E_J/E_C$ at
$E_J /E_C \leq 1$ if the ohmic resistance originates from
quasiparticles ($R \sim R_{qp}$) \cite{SZ}. However, all our experimental data
for small $E_J /E_C$ were obtained for shunted junctions ($R \sim R_s$), so we
ignore this deviation from the vertical transition line.
\bibitem{SIT} In the past, the difference between DPT
and SIT was ignored, the latter term being used more frequently. In order to
emphasize the link with previous studies, the term SIT was used 
with quotation marks in the 
title of our Letter although this paper
deals mostly with DPT.
\end{references}
\end{document}